\documentstyle[12pt,epsf]{article}
\advance \textheight \topskip

\begin{document}
\begin{flushright}
preprint hep-ph96xxxxx\\
Preprint DFPD 96/TH/61
\end{flushright}
\centerline{\Large {\bf The Paths of Unification In The GUST}}
\centerline{\Large {\bf With The $G \times G$ Gauge Groups of
$E(8) \times E(8)$ }}

\vspace{24pt}
\begin{center}
{\Large {\bf A. Maslikov $^{b}$ , I. Naumov $^{b}$, G. Volkov $^{a,b}$ }}

\bigskip
$^a$ INFN Sezione di Padova and Dipartimento di Fisica \\
Universit\`a di Padova, \\
Via Marzolo 8, 35100 Padua, Italy \\

\bigskip
$^b$ Institute for  High Energy Physics\\
142284 Protvino, Moscow Region, Russia\\
\end{center}

\bigskip
\centerline{ABSTRACT}
In the framework of the four dimensional heterotic superstring with free
fermions we discuss the rank eight and/or sixteen
Grand Unified String Theories (GUST) which contain
the $SU(3)_H$-gauge family symmetry.
We explicitly investigate the paths of the unification
in the  GUST with gauge symmetry
$ G\times G = [SU(5) \times U(1)\times (SU(3) \times U(1))_H]^{\otimes 2}$.
We show that the GUSTs with the $G\times G$ gauge group allow to make
the scale of unification to be consistent with
the string scale $M_{SU} \sim g_{string}\cdot 5 \cdot 10^{17} GeV$.

\newpage

\section{Introduction}

For a couple of years superstring theories, and particularly the heterotic
string theory \cite {1',2'}, have provided an efficient way to construct the
Grand Unified
Superstring Theories ($GUST$) of all known interactions, despite the
fact that it is
still difficult to construct unique and fully realistic low energy models
resulting after decoupling of massive string modes.

In the fermionic formulation of the four-dimensional heterotic string theory
\cite {3',4'} in
addition to the two transverse bosonic coordinates $X_{\mu}$ ,$\bar{X}_{\mu}$
and their left-moving superpartners ${\psi}_{\mu}$, the internal sector
${\cal M}_{c_L;c_R}$ contains 44 right-moving ($c_R=22$) and 18 left-moving
($c_L=9$) real fermions (each real world-sheet fermion has
$c_f=1/2$).
String theories possess infinite dimensional symmetries that place many
specific constraints on the theory spectrum. These symmetries origin from
2 dimensional conformal invariance, modular invariance, and Virasoro and
Kac-Moody algebras.
Because of the presence of the affine Kac-Moody algebra (KMA)
$\hat g$ (which is a 2-dimensional manifestation of gauge symmetries of the
string itself) on the world sheet, string constructions yield definite
predictions concerning representation of the symmetry group,
especially for the rank 8 and greater,  that can be used
for low energy models building.

There are not so many GUSTs describing the observable sector of Standard
Model. They are well known: the $SU(3)\times SU(2)\times U(1)^n\times G_{hid}$
gauge group, the Pati-Salam ($SU(4)\times SU(2)\times SU(2)\times U(1)
\times G_{hid}$) gauge group,
the flipped $SU(5) \times U(1)\times G_{hid}$ gauge group
and $SO(10)\times G_{hid}$ gauge group \cite{20',20''}.
For the heterotic 10-dimensional  string the groups $E_8\otimes E_8$ and
$spin(32)/Z_2$ are characteristic. Hence it is interesting to consider GUSTs
in four dimension based on its various rank 8 and 16 subgroup \cite{25'}.
As the GUSTs originating from level one Kac-Moody algebra
(KMA) contain only low-dimensional representations, new types of GUSTs with
the $G \times G$ gauge groups can naturally
appear in consideration \cite {25',26'}.
Moreover for the observable gauge symmetry one can consider the diagonal
subgroups ${G'}^{sym}$ of the rank 16 group $G \times G \subset SO(16)
\times SO(16)$ or $\subset E(8) \times E(8)$.
Early \cite {25'} we considered the possible ways of breaking the "string"
gauge subgroups
$\subset E_8\otimes E_8$ down to low energy supersymmetric model that includes
Standard Model group and horizontal factor $SU(3)$.
There are good physical reasons for including the horizontal $SU(3)_H$ group
into the unification scheme. Firstly, this group naturally accommodates three
fermion families presently observed (explaining their origin) and, secondly,
can help to solve the flavour problem in SUSY GUTs and
can provide correct and economical description of the fermion mass spectrum
and mixing  without invoking high dimensional representation of conventional
$SU(5)$, $SO(10)$ or $E(6)$ gauge groups\cite{9'}.
Construction of a string model
(GUST) containing the horizontal gauge symmetry provides additional strong
motivation to this idea. Moreover, the fact that in GUSTs high dimensional
representations are forbidden by the KMA is a very welcome feature in
this context.
The constraints of horizontal model parameters followed from this approach
allow the existence of the interesting flavour-changing physics in the
TeV region. Also these models gives rise to a rather natural way of
the superweak-like CP-violation\cite {9'}.
All this leads us naturally to considering possible forms for horizontal
symmetry
$G_H$, and $G_H$ quantum number assignments for quarks (anti-quarks) and
leptons (anti-leptons) which can be realized within GUST's framework.

Here we present shortly the string models including Grand Unification group
$[SU(5)\times U(1) \times G_H]^{\otimes 2}$,
along with horizontal gauge symmetry $G_H=U(3)$.
Using the (2,0) world-sheet
 superconformal symmetry  we study the superpotential.
The form of obtained superpotential implies that 2
generations remain massless comparing with the $M_W$ scale.
Using the condition of  $SU(3)_H$ anomaly cancellation  the theory
predicts the existence of the Standard Model singlet
"sterile" particles that participate only in horizontal $SU(3)_H$
and/or $U(1)_H$
interactions.  As following from the form of the superpotential some of
them   could be light (much less than $M_W$) that will be very interesting in
sense of experimental accelerator and astrophysical searches.
In this model after anomalous $U(1)$ D-term  suppressing  the only surviving
horizontal gauge group is $SU(3)_H$.

We outline the perspective way of  including the symmetric subgroup on the
intermediate stage that does not involve higher level of Kac-Moody
algebra representations.
Starting from the rank 16 grand unified gauge group
of the form $G\times G$
\cite{25',26'} and making use of the KMA which select
the possible gauge group
representations we
discuss some  ways of breaking of string rank 16 gauge
group $[SU(5)\times U(1) \times G_H]^{\otimes 2}$
down to the  symmetric diagonal subgroups \cite {25'},\cite {Barbieri'}.
This model allows two ways of embedding
chiral matter (16 quarks, leptons and right neutrino) in  $\underline {1}$,
 $\underline {\bar 5}$
and $\underline {10}$ representation of $SU(5) \times U(1)$, which
 correspond to the flipped and non-flipped $SU(5) \times U(1)$
models respectively \cite {Barr'}.

 The main goal of our paper is to solve the problem of discrepancy between
the unification scales of
$SU(3^c)$, $SU(2)_{EW}$, $U(1)_Y$-gauge coupling constants ,
$M_G \sim 10^{16} GeV$,
and string scale in GUSTs, $M_{GUST}= M_{SU}=g_{string}\cdot 5\cdot 10^{17}
GeV$.

We consider two possibilities of the breaking of the primordial
$[U(5) \times G_H]^{\otimes 2}$ gauge symmetry with two variants of
$Q_{em}$ charge  quantization correspondingly.
For the various chains of gauge symmetry breaking in flipped and non-flipped
 cases of the $SU(5)$ model we carry out the RGE analysis of the behaviour of
the gauge coupling constants taking into account the  possible intermediate
thresholds (the additional Higgs doublets, color triplets,
SUSY threshold, massive fourth generation)
 and the threshold effects due to the  massive string states.
We show that only in
non-flipped case of $[U(5) \times G_H]^{\otimes 2}$ GUST it is possible
to make the unification scale of $g_{1,2,3}$-
coupling constants in supersymmetric standard model, $M_G$, to be
consistent  with the string
scale unification, $M_{GUST} = M_{SU}$ and
obtain estimation of the string coupling constant $g_{str}=O(1)$.
As an additional benefit,
the values of the $g_{str}$ and $M_{SU}$ allow us to estimate
the horizontal coupling constant $g_{3H}$ on the scale of $\sim$1 TeV.

\section{The features of the GUST spectrum with
$[SU(5)\times U(1)\times SU(3)\times U(1)]^{\otimes 2}.$
The ways of the gauge symmetry breaking.}\label{sbsec42}

Model~1 is defined by 6 basis vectors given in Table \ref{tabl1} which
generates the $Z_2\times Z_4\times Z_2\times Z_2\times Z_8\times Z_2$
group under addition.

\begin{table}[h]
\caption{\bf Basis of the boundary conditions for all world-sheet fermions.
Model~1.}
\label{tabl1}
\footnotesize
\begin{center}
\begin{tabular}{|c||c|ccc||c|cc|}
\hline
Vectors &${\psi}_{1,2} $ & ${\chi}_{1,..,6}$ & ${y}_{1,...,6}$ &
${\omega}_{1,...,6}$
& ${\bar \varphi}_{1,...,12}$ &
${\Psi}_{1,...,8} $ &
${\Phi}_{1,...,8}$ \\ \hline
\hline
$b_1$ & $1 1 $ & $1 1 1 1 1 1$ &
$1 1 1 1 1 1 $ & $1 1 1 1 1 1$ &  $1^{12} $ & $1^8 $ & $1^8$ \\
$b_2$ & $1 1$ & $1 1 1 1 1 1$ &
$0 0 0 0 0 0$ &  $0 0 0 0 0 0 $ &  $ 0^{12} $ &
${1/2}^8 $ & $0^8$  \\
$b_3$ & $1 1$ & $ 1 1 1 1 0 0 $ & $0 0 0 0 1 1 $ & $ 0 0 0 0 0 0 $ &
$0^4  1^8 $ & $ 0^8 $ &  $ 1^8$ \\
$b_4 = S $ & $1 1$ &
$1 1 0 0 0 0 $ & $ 0 0 1 1 0 0 $ & $ 0 0 0 0 1 1 $ &
$ 0^{12} $ & $ 0^8 $ & $ 0^8 $ \\
$  b_5 $ & $ 1 1 $ & $ 0 0  1 1 0 0 $ &
$0 0 0 0 0 0 $ &  $1 1  0 0 1 1$ &  $ 1^{12} $ &
${ 1/4}^5  {-3/4}^3 $ & $ {-1/4}^5\ {3/4}^3  $  \\
$  b_6 $ & $ 1 1 $ & $ 1 1  0 0  0 0 $ &
$ 0 0  0 0 1 1 $ &  $0 0  1 1 0 0$ &  $ 1^2 0^4 1^6 $ &
$ 1^8  $ & $ 0^8  $  \\
\hline \hline
\end {tabular}
\end{center}
\normalsize
\end{table}

The  model corresponds to the following chain of the gauge
symmetry breaking:
$\longrightarrow U(8)^2
\longrightarrow [U(5)\times U(3)]^2\ . $

Since the matter fields form the chiral multiplets of $SO(10)$, it is possible
to write down  $U(1)_{Y_5}$--hypercharges of massless states. In order to
construct the correct electromagnetic charges for matter fields we must define
the hypercharges operators for the observable $U(8)^{I}$ group as follows
\begin{equation}\label{eq16}
Y_5=\int^\pi_0 d\sigma\sum_a \Psi^{*a}\Psi^a ,\,\,\,\,\,
Y_3=\int^\pi_0 d\sigma\sum_i \Psi^{*i}\Psi^i
\end{equation}
and analogously for the $U(8)^{II}$ group.

Then the orthogonal combinations
\begin{equation}\label{eq17}
 \tilde Y_5 = {1\over 4}(Y_5 + 5Y_3), \,\,\,\,\,
 \tilde Y_3 = {1\over 4}(Y_3 - 3Y_5),
\end{equation}
play the role of the hypercharge operators of $U(1)_{Y_5}$ and
$U(1)_{Y_H}$ groups
respectively. In  Table \ref{tabl3'} we give the hypercharges
 $\tilde Y_5^{I},\tilde Y_3^{I}, \tilde Y_5^{II},\tilde Y_3^{II}$.

\begin{table}[t]
\caption{\bf The list of quantum numbers of the states. Model~1.}
\label{tabl3'}
\footnotesize
\noindent \begin{tabular}{|c|c||c|cccc|cccc|} \hline
N$^o$ &$ b_1 , b_2 , b_3 , b_4 , b_5 , b_6 $&
$ SO_{hid}$&$ U(5)^I $&$ U(3)^I $&$ U(5)^{II} $&$
U(3)^{II} $&$ {\tilde Y}_5^I $&$ {\tilde Y}_3^I $&$ {\tilde Y}_5^{II} $&$
{\tilde Y}_3^{II}$ \\ \hline \hline
1 & RNS &&5&$\bar 3$&1&1&--1&--1&0&0 \\
  &     &&1&1&5&$\bar 3$&0&0&--1&--1 \\
  &0\ 2\ 0\ 1\ 2(6)\ 0&&5&1&5&1&--1&0&--1&0 \\
${\hat \Phi }$
  &&&1&3&1&3&0&1&0&1 \\
  &&&5&1&1&3&--1&0&0&1 \\
  &&&1&3&5&1&0&1&--1&0 \\ \hline \hline
2 &0\ 1\ 0\ 0\ 0\ 0&&1&3&1&1&5/2&--1/2&0&0 \\
  &&&$\bar 5$&3&1&1&--3/2&--1/2&0&0 \\
  &&&10&1&1&1&1/2&3/2&0&0 \\
${\hat \Psi }$
  &0\ 3\ 0\ 0\ 0\ 0&&1&1&1&1&5/2&3/2&0&0 \\
  &&&$\bar 5$&1&1&1&--3/2&3/2&0&0 \\
  &&&10&3&1&1&1/2&--1/2&0&0 \\ \hline
3 &0\ 0\ 1\ 1\ 3\ 0&$-_1\ \pm_2$&1&1&1&3&0&--3/2&0&--1/2 \\
  &0\ 0\ 1\ 1\ 7\ 0&$-_1\ \pm_2$&1&$\bar 3$&1&1&0&1/2&0&3/2 \\
${\hat \Psi }^H$
  &0\ 2\ 1\ 1\ 3\ 0&$+_1\ \pm_2$&1&$\bar 3$&1&3&0&1/2&0&--1/2 \\
  &0\ 2\ 1\ 1\ 7\ 0&$+_1\ \pm_2$&1&1&1&1&0&--3/2&0& 3/2 \\ \hline
4 &1\ 1\ 1\ 0\ 1\ 1&$\mp_1\ \pm_3$&1&1&1&$\bar 3$&0&--3/2&0&1/2 \\
  &1\ 1\ 1\ 0\ 5\ 1&$\mp_1\ \pm_3$&1&$\bar 3$&1&1&0&1/2&0&--3/2 \\
${\hat \Phi }^H$
  &1\ 3\ 1\ 0\ 1\ 1&$\pm_1\ \pm_3$&1&$\bar 3$&1&$\bar 3$&0&1/2&0&1/2 \\
  &1\ 3\ 1\ 0\ 5\ 1&$\pm_1\ \pm_3$&1&1&1&1&0&--3/2&0&--3/2 \\ \hline
5 &0\ 1(3)\ 1\ 0\ 2(6)\ 1&$-_1\ \pm_3$&1&3($\bar 3$)&1&1&$\pm$5/4&$\pm$1/4
&$\pm$5/4&$\mp$3/4 \\
  &&$+_1\ \pm_3$&5($\bar 5$)&1&1&1&$\pm$1/4&$\mp$3/4&$\pm$5/4&$\mp$3/4 \\
${\hat \phi }$
  &0\ 1(3)\ 1\ 0\ 4\ 1&$-_1\ \pm_3$&1&1&1&3($\bar 3$)&$\pm$5/4&$\mp$3/4
&$\pm$5/4&$\pm$1/4 \\
  &&$+_1\ \pm_3$&1&1&5($\bar 5$)&1&$\pm$5/4&$\mp$3/4&$\pm$1/4&$\mp$3/4 \\
\hline
6 &1\ 2\ 0\ 0\ 3(5)\ 1&$\pm_1\ -_4$&1&1&1&1&$\pm$5/4&$\pm$3/4
&$\mp$5/4&$\mp$3/4 \\
  &1\ 1(3)\ 0\ 1\ 5(3)\ 1&$+_1\ \mp_4$&1&1&1&1&$\pm$5/4&$\pm$3/4
&$\pm$5/4&$\pm$3/4 \\
${\hat \sigma }$
  &0\ 0\ 1\ 0\ 2(6)\ 0&$\mp_3\ +_4$&1&1&1&1&$\pm$5/4&$\mp$3/4
&$\pm$5/4&$\mp$3/4 \\ \hline
\end{tabular}
\normalsize
\end{table}

With the chiral matter and "horizontal" Higgs fields available in Model~1
 the possible form of the renormalizable (trilinear)
part of the superpotential responsible for fermion mass matrices is
restricted not only by  the gauge symmetry.
Another strong constraint
 comes from the interesting observation that a modular invariant
N=1 space-time supersymmetric theory also extends to a global N=2
world sheet superconformal symmetry \cite{bank'} which now contains
 two distinct fermionic components of the energy-momentum tensor,
 $T_F^+$ and  $T_F^-$
 and there is also the  $U_J(1)$ current $J$. This conserved $U(1)$ current of
the N=2 superalgebra may play a key role in
constructing of realistic phenomenology. Thus all vertex operators
have the definite $U(1)$ charge.
Let us consider the contribution $R^2(NS)$ to the   three point
fermion-fermion-boson matter superpotential:

\begin{eqnarray}\label{eq32}
W_1&=& g\sqrt{2} \biggl[ {\hat \Psi}_{(1,3)}
{\hat \Psi}_{({\bar 5},1)}
{\hat\Phi}_{(5,{\bar 3})} +
 {\hat \Psi}_{(1,1)} {\hat \Psi}_{({\bar 5},3)}
{\hat\Phi}_{(5,{\bar 3})} + \nonumber\\
&+& {\hat \Psi}_{(10,3)} {\hat \Psi}_{({\bar 5},3)}
{\hat\Phi}_{({\bar5},3)} +
{\hat \Psi}_{(10,3)} {\hat \Psi}_{(10,1)}
{\hat\Phi}_{(5,{\bar 3})} \biggr]
\end{eqnarray}

From the above form of the Yukawa couplings it
follows that two (chiral) generations
have to be very light (comparing to $M_W$ scale).

The $SU(3)_H$ anomalies of the matter fields (row No 2) are
naturally canceled by
the chiral "horizontal" superfields  forming two sets:
${\hat \Psi}^H_{(1,N;1,N)}$ and ${\hat \Phi}^H_{(1, N;1, N)}$,
{}~$\Gamma = \underline 1$,~~$N = \underline 1, \,  \underline 3$,~
(with both ${SO(2)}$ chiralities, see Table \ref{tabl3'}, row No 3, 4
respectively).
The superpotential, $W_2$, consists of the following $R^2 NS$-terms:
\begin{eqnarray}\label{eq34}
W_2&=& g\sqrt{2} \biggl[
{\hat \Phi}^H_{(1,1;1,\bar3)}{\hat \Phi}^H_{(1,\bar 3;1,1)}
{\hat \Phi}_{(1,3;1,3)} +
{\hat \Phi}^H_{(1,1;1,1)}{\hat \Phi}^H_{(1,\bar 3;1,\bar 3)}
{\hat \Phi}_{(1,3;1,3)} + \nonumber\\
&+& {\hat \Phi}^H_{(1,\bar 3;1,\bar 3)}{\hat \Phi}^H_{(1,\bar 3;1,\bar 3)}
{\hat \Phi}_{(1,\bar 3;1,\bar 3)} +
 {\hat \Psi}^H_{(1,{\bar 3};1,1)}{\hat \Psi}^H_{(1,{\bar 3};1, 3)}
{\hat \Phi}_{(1,{\bar 3};1,{\bar 3})} +\nonumber\\
&+&{\hat \Psi}^H_{(1,1;1,3)}
{\hat \Psi}^H_{(1,{\bar 3};1,3)}
{\hat \Phi}_{(1, 3;1, 3)}
+{\hat \sigma }_{(-_1 -_4 )} {\hat \sigma }_{(+_1 +_4 )}
{\hat \Psi }_{(1,1;1,1)} \biggr]
\end{eqnarray}

From (\ref{eq34}) it follows that some of the horizontal fields in sectors
(No 3, 4) remain massless at the tree-level.
This is a remarkable prediction: "horizontal" fields are "sterile",
e.g. they interact with the ordinary
chiral
matter fields only through the $U(1)_H$ and $SU(3)_H$
gauge boson
and therefore this "sterile" matter
is of an  interest in the context of the
experimental searches on accelerators or in astrophysics.
The Higgs fields  could give the following
$(NS)^3$ contributions to the renormalizable superpotential:
\begin{eqnarray}\label{w3}
W_3=\sqrt{2}g\biggl\{
{\hat\Phi }_{ (5,1;1,3)}  {\hat\Phi }_{ (\bar 5,1;\bar 5,1)}
{\hat\Phi }_{ (1,1;5,\bar 3)} +
{\hat\Phi }_{ (5,1;1,3)}  {\hat\Phi }_{ (1,\bar 3;1,\bar 3)}
{\hat\Phi }_{ (\bar 5,3;1,1)} && \nonumber\\
+ {\hat\Phi }_{ (1,3;5,1)}  {\hat\Phi }_{ (\bar 5,1;\bar 5,1)}
{\hat\Phi }_{ (5,\bar 3;1,1)} +
{\hat\Phi }_{ (1,3;5,1)}  {\hat\Phi }_{ (1,\bar 3;1,\bar 3)}
{\hat\Phi }_{ (1,1;\bar 5,3)} +\mbox{conj.}
\biggr\}&&
\end{eqnarray}

So, $W_1 +W_2 +W_3$ is the most general renormalizable superpotential
which includes all nonzero three-point (F-type) expectation values
of the vertex operators for corresponding 2-dimensional conformal
model.
The only non-vanishing nonrenormalizable superpotential $W_4=R^4$
is as follows
\begin{eqnarray}
W_4\sim \frac{g^2}{M_{Pl}}{\hat\Psi}_{(1,3;1,1)}
{\hat\Phi}^H_{ (+_1,-_3)(1,\bar 3;1,1)}
{\hat\sigma}_{1\ (-_1,-_4)} {\hat\sigma}_{3\ (+_3,+_4)}.
\end{eqnarray}

As we can see from the states list for Model 1, the hidden group
$U(1)_1$ in this model appears to be anomalous: Tr $U(1)_1=12$.

This means that at one-loop string level there exists Fayet-Iliopoulos
D-term determined by VEV of the dilaton and it is proportional to
Tr $U(1)_1=12$.
Potentially this term could break SUSY at the high scale
and destabilize the vacua \cite{Dine'}. This could be avoided
if the potential has
D-flat direction on $U(1)_1$-charged fields which have VEVs that break
anomalous group (and may be some other groups), compensate D-term, and
restore SUSY. Those fields have to have appropriate charges on the
remaining groups in order that  cause no  SUSY breaking via their D-terms.
In our case we can avoid the SUSY breaking caused by D-term by using the
pairs of fields ${\hat\phi}_1$ and/or ${\hat\phi}_3$ from the sector 5.
In addition to anomalous group breaking we also obtain breaking of the
groups $U(1)_3^{hid}$ (hidden),
$U(1)^I_H\times U(1)^{II}_H\rightarrow U(1)_H^{'} $
(in the case of using only one of ${\hat\phi}_1$, ${\hat\phi}_3$,
otherwise the group is broken totally)
and
$U(1)^I_5\times
U(1)^{II}_5\rightarrow U(1)^{sym}_5$, and also horizontal
group $SU(3)_H^{I}\rightarrow SU(2)_H^I$ and/or
$SU(3)_H^{II}\rightarrow SU(2)_H^{II}$. For complete breaking
the initial horizontal symmetry $SU(3)_H^I \times SU(3)_H^{II}$
we also need to use the VEVs of the ${\hat{\Phi}}_{(1,3;1,3)}$- Higgs fields.
Besides that fields potentially the fields ${\hat\sigma}_2$
can also obtain VEVs,
that  additionally breaks down
the hidden group $SO(6)_4$ and also breaks  $U(1)_H^{'}$.
Note that
since the  superfields mentioned above besides the $\hat{\sigma}_2$-field
do not participate in construction
of the renormalizable superpotential $W_1+W_2+W_3$
in this scheme we have no problem with the F-flat directions.
Finally, note that in non-flipped $SU(5)\times U(1)$ model
we can give the VEVs to the fields ${\hat\sigma}_1$
for breaking $U(1)^{sym}_5$. In this case for the choice of the $D$-flat
direction we also need to use VEVs of the fields ${\hat\sigma}_3$.
From the form of the ${\hat\sigma}$-depended contribution to the
superpotential $W_2$  (\ref{eq34}) it follows that the field
of the fourth generation ${\hat\Psi}_{(1,1;1,1)}$ (the fourth neutrino
in non-flipped scheme)  obtains a heavy mass.

 Further we shortly discuss the problem of
  gauge symmetry breaking in Model~1.  The most
important point  is that the Higgs fields
$(\underline{10}_{1/2}+\underline{\bar{10}}_{-1/2})$ do not appear.
However there exists some
possibilities to break the GUST group $[(U(5)\times U(3))^I]^{\times 2}$
down to the symmetric subgroups using the following VEVs of the Higgs fields
$(\underline 5,\underline 1;\underline 5,\underline 1)_{(-1,0;-1,0)}$:
\begin{description}
\itemsep=0cm
\item[a)] $\qquad\qquad <(\underline 5,\underline 5)> =
a\cdot \mbox{diag}(1,1,1,1,1)$
\item[b)] $\qquad\qquad <(\underline 5,\underline 5)>_0 =
 \mbox{diag}(x,x,x,y,y)$,
\item[c)] $\qquad\qquad <(\underline 5,\underline 5)>_0 =
a \cdot \mbox{diag}(0,0,0,1,1)$,
\item[d)] $\qquad\qquad <(\underline 5,\underline 5)>_0 =
a \cdot \mbox{diag}(1,1,1,0,0)$.
\end{description}
With the VEV a) the GUST group $[(U(5)\times U(3))^I]^{\otimes 2}$
 breaks to symmetric group:
$$ \mbox{\bf a)}\qquad\qquad U(5)^I\times U(5)^{II} \rightarrow {U(5)}^{sym}
\rightarrow ...$$
With such a breaking tensor Higgs fields transform
under the $(SU(5)\times U(1))^{sym}\times G_H$ group in the
following way:
\begin{eqnarray}\label{eq19}
(\underline 5,\underline 1;\underline 5,\underline 1)_{(-1,0;-1,0)}
{}~&\rightarrow~~& {(\underline {24},\underline 1) }_{(0,0)}~~
+~~(\underline 1,\underline 1) _{0,0}.
\end{eqnarray}

The diagonal VEVs
of the Higgs fields break the GUST with $G \times G$ down to
the "skew"-symmetric group with the generators $\triangle^{sym}$ of the form:
\begin{equation}\label{eq21}
\triangle^{sym} (t) = -t^* \times 1 + 1\times t,
\end{equation}
The corresponding  hypercharge of the symmetric group reads:
\begin{equation}\label{eq22}
\bar Y = \tilde {Y}^{II} - \tilde {Y}^{I}.
\end{equation}
Adjoint  representations which appear on the $rhs$ of (\ref{eq19}) can be used
for
further breaking of the symmetric group. This can lead to the final physical
symmetry
$$ \qquad SU(3^c)\times SU(2_{EW})\times U(1)_5 \times U(1)^{sym}
\times G_H^{'}$$
with the low-energy gauge symmetry of the quark--lepton generations with an
additional $U(1)^{sym}$--factor. As we already discussed above
the form of the $G_H^{'}$ depends on the way of the U(1)
anomaly group cancellation and the complete breaking of this group is
realized by the VEVs of the ${\hat \Phi}_{(1,3;1,3)}$ Higgs fields
and/or ${\hat \Phi}_{(1,1;5,{\bar 3})}$,
${\hat \Phi}_{(5,{\bar 3};1,1)}$ Higgs fields.

Note, that when we use the VEVs b),c),d) there exist also
the others interesting ways of breaking the $G^I \times G^{II}$ gauge symmetry
down to
\begin{description}
\itemsep=0cm
\leftmargin=5mm
\rightmargin=0mm
\item[b)] $\quad SU(3^c) \times
 SU(2)_{EW} \times U(1)_5 \times U(1)^{sym} \times G_H^{'} \rightarrow \ldots$,
\item[c)] $\quad  SU(3^c) \times SU(2)^I_{EW}
\times SU(2)^{II}_{EW} \times U(1)_5\times U(1)^{sym}
\times G_H^{'} \rightarrow \ldots $ ,
\item[d)] $\quad SU(3^c)_I \times SU(3^c)_{II} \times SU(2)_{EW} \times U(1)_5
\times U(1)^{sym} \times G_H^{'} \rightarrow \ldots$ .
\end{description}
 We could consider the GUST construction involving
$[SO(10)  \times G_H]^{\otimes 2}\times G_{hidden}$
 as the  gauge group \cite{25'}.
In that model the only Higgs fields appeared are $(10,10)$ of
$SO(10) \times SO(10) $ and the hidden group $G_{hidden}=U(1)\times SO(6)$
 is anomaly free.
As an illustration we would like to remark that for the
$SO(10) \times SO(10) \times G_H \times G_H$ GUST we can consider similarly
the following VEVs of the Higgs fields $(10,10)$:
\begin{description}
\itemsep=0cm
\item[a')]$\quad <(10,10)>_0 = a\cdot \mbox{diag} (1,1,1,1,1,1,1,1,1,1)$,
\item[b')]$\quad <(10,10)>_0 = a \cdot \mbox{diag} (1,1,1,1,1,1,0,0,0,0)$,
\item[c')]$\quad <(10,10)>_0 = a \cdot \mbox{diag} (1,1,1,1,1,1,x,x,x,x)$.
\end{description}
These cases
lead  correspondingly to the following chains of
$[SO(10)]^{\otimes 2}$ breaking ways:
\begin{description}
\itemsep=0cm
\item[a')]$\quad [SO(10)]^{sym}$,
\item[b')]$\quad SU(4)\times SU(2)_{I1}\times SU(2)_{I2}\times
 SU(2)_{II1} \times SU(2)_{II2}$,
\item[c')]$\quad SU(4)\times SU(2)_{I}\times SU(2)_{II}$.
\end{description}

\section{The GUT and string unification scale. The paths of unification in
flipped and non-flipped GUST models.}

Indeed the estimates on the $M_{H_0}$ scale depend on the
value of the family gauge coupling.
String theories imply a natural unification of the gauge
and gravitational
couplings, $g_i$ and $G_N$ respectively. For example, it turns out that
these couplings unify
at tree level to form one coupling constant $g_{string}$ \cite{Ginsparg'}:
\begin{eqnarray}
8\pi \frac{G_N}{\alpha'}=g_ik_i=g^2_{str}.
\end{eqnarray}
\medskip
Here $\alpha'$ is the Regge slope, the coupling constants $g_i$
correspond to the gauge group $G_i$ with the Kac-Moody levels $k_i$.
In string theory the scale of unification is fixed by the Planck scale
$M_{Pl}\approx 10^{19}$.
In one-loop string calculations
the value of the unification scale could be divided into moduli independent
part and a part that depends on the VEVs of the moduli fields. The latter
part considered as the correction to the former and obviously it is
different for the various models. Like in the gauge fields theory
this correction is called the string threshold correction of the massive
string states. Moduli independent contribution depends on the renormalizing
scheme used,
so in $\overline{DR}$ renormalization scheme the scale of string
unification is shifted to \cite{Kaplunovsky'}a :
\begin{equation}
M_{SU} = \frac{e^{(1-\gamma)/2} 3^{-3/4}}{4\pi} g_{str} M_{Pl}
 \approx 5\,g_{str} \times {10}^{17} GeV. \label{Mstring}
\end{equation}

There  exists the most important difference between the unification
scales of gauge coupling
unification in string theory and  in field theory. In field theory this scale
is determined via extrapolation of data within the Supersymmetric
$SU(3) \times SU(2) \times U(1)$ Standard Model using
 the renormalization group equations (RGE) for the gauge couplings
$${\alpha}_{3}^{-1}(M_Z) = 8.93\ ,$$
$${\alpha}_{2}^{-1}= ({\alpha}_{em}/
{\sin}^{2}{\theta}_{W})^{-1}|_{M_Z}=29.609\ ,$$
$${\alpha}_{1}^{-1}= (5{\alpha}_{em}/ 3{\cos}^{2}
{\theta}_{W})^{-1}|_{M_Z}=58.975\ .$$
The factor 5/3 in the definition of ${\alpha}_1$ has been included for the
normalization at the unification scale $M_{G}$.

The one loop   renormalization group equations for these gauge couplings
 are given by

\begin{equation}
\frac {d{\alpha}_{i}}{d \ln \mu}= \frac{{\alpha}_{i}^2}{2\pi} b_i
\label{RGEi}
\end{equation}

 Beta functions coefficients
for the $SU(3)\times SU(2) \times U(1)$ coupling constants
in SUSY models are given by the following:

\begin{eqnarray}
b_3&=& -9 +2N_g + 0\cdot N_h + \frac{1}{2}\cdot N_3,\\
b_2&=& -6 +2N_g + \frac{1}{2}\cdot N_h +  0\cdot N_3,\\
b_1&=&\,0 +2N_g + \frac{3}{10}\cdot N_h + \frac{1}{5}\cdot N_3,
\label{bi}
\end{eqnarray}
where $N_g$ is the number of generation and $N_h$ is the number of Higgs
doublets and we also include some possible intermediate thresholds for heavy
Higgs doublets $(N_h - 2)$ and color triplets, $N_3$,
which exist in the spectrum of Model 1 and can be take into account for
RGE.

The RGE are integrated from $M_Z$-mass to the unification scale $M_{G}$.
In the presence of various intermediate scale, $M_I$, $I=1,2,3,...$,
the RGE are given by:

\begin{eqnarray}
{\alpha}_i^{-1}(M_Z)=
{\alpha}_i^{-1}(M_{G}) + \frac{b_i}{2\pi}\ln{\frac{M_{G}}{M_Z}}
-\frac{b_{i1}}{2\pi}\ln{\frac{M_1}{M_Z}}
-\frac{b_{i2}}{2\pi}\ln{\frac{M_2}{M_Z}}-...
\end{eqnarray}

where $b_{iI}$ are the additional corresponding contributions of the new
thresholds to the beta functions. At the Z-mass
scale we have:

\begin{eqnarray}
{\sin}^2{\theta}_{W}(M_Z) &\approx & 0.2315 \pm 0.001\ ,\nonumber\\
M_Z&=&91.161\pm 0.031 GeV, \label{experiment}
\end{eqnarray}

Note, that for flipped models
\begin{equation}
\sin^2 \theta_W(M_{G}) =\frac{15k^2}{16k^2 +24}
\end{equation}
where $k^2=g_1^2/g_5^2$.
In the SO(10) limit (or for non-flipped case) we have
$k^2=1$ and $\sin^2 \theta_W = 3/8$.

The string unification scale could be contrasted with
MSSM, $SU(3^c)\times SU(2) \times U(1)$
naive unification scale,
\begin{equation}
M_{G} \approx 2\times {10}^{16} GeV
\end{equation}
obtained by running the SM particles and their SUSY-partners to high energies.

One of the first way to explain the difference between these two mass
scales, $M_{SU}$ and $M_{MSSM}=M_G$, was the  attempts to take into account
the  string thresholds corrections of the massive string states
\cite {Kaplunovsky'}:
\begin{equation}
\frac{1}{g_i^2(\mu)} = k_i\frac{1}{g_{string}^2}
+2b_i\ln(\frac{\mu}{ M_{SU}}) + {\tilde\Delta}_{G_i}.
\end{equation}
where the index $i$ runs over gauge coupling and  $\mu$ is
some phenomenological scale such as $M_Z$ or $M_{G}$.
The coefficients $k_i$ are the Kac-Moody levels (e.g.
for SU(5) $k_2=k_3=1,k_1=5/3$).
The quantities $ {\tilde\Delta}_{G_i}$ represent
the heavy thresholds corrections, which are the corrections arising
from the infinite towers of massive string states.
Although these states have the Planck mass scale, there are infinite
number of them, so they together could have the considerable effect.
In general the full string thresholds corrections are  of the form
\begin{equation}
{\tilde {\Delta}}_i={\Delta}_i +k_i Y,
\end{equation}
where Y is independent of the gauge
group factor. Moreover, the low energy predictions for
${\sin}^2{\theta}_W(M_Z)$ and ${\alpha}_3(M_Z)$ depend only on the differences
$
({\tilde{\Delta}}_i -{\tilde{\Delta}}_j)=(\Delta_i -\Delta_j)
$
for the different gauge groups.

However the $Y$ factor makes influence on the estimation of the value of
$g_{str}$ if we base on low energy gauge constants and use RGE. Note that
in general $g_{str}$ is defined by VEV of the dilaton moduli field but
because of degeneracy of the classic potential of the moduli fields
we do not know the $g_{str}$ a priori. If we could have the value of the
$g_{str}$ then the its coincidence with our estimates will show the
correctness of our model.

It is supposed that the value of the $Y$ factor is small enough
\cite{KalaraLopez} so we neglect it in our calculations of $g_{str}$
via RGE.

In the  GUSTs examples considered the threshold corrections
of the massive string states are not large enough to
compensate the difference between the scales of unification of
the string and the MSSM (GUT) models.
Later we will discuss the possible effects of them in $G \times G$ models,
for example in Model 1.

In our calculations for Model 1
we will follow the way suggested in \cite{Kaplunovsky'}c,\cite {Dienes'}.
According to it
we have to calculate an integral of the modified partition function
over the fundamental domain $\Gamma$ of modular group.
$$
\delta\Delta=\int_\Gamma\frac{\mbox{d}^2\tau}{\tau_2}
\left(|\eta(\tau)|^{-4}\hat{Z}(\tau)-b_G\right)
$$
where $b_G$ is the beta function coefficient,
$\hat Z(\tau)$ is modified partition function.  Modified in this
context means that the charge operators $Q$ are inserted in the trace
in partition function in the following way:
$$ \hat
Z(\tau)=-\mbox{Tr}Q^2_sQ^2_Iq^Hq^{\bar H}, $$
where $Q_s$ is helicity operator and
$Q_i$ is a generator of a gauge group.  We are interesting in particular
in the difference between groups $SU(5)_I$ and $U(1)_I$ in Model~1.
For this case charge polynomial is $5Q_1Q_2$  (see \cite{Kaplunovsky'}c).
Rewritten via well known theta
function, the modified function $\hat Z(\tau)$ for our case reads:
$$ \hat{Z}(\tau)=-\frac5{512}\sum_{{\bf a,b}}c\left({\bf a\atop b}\right)
\eta^{-1}\theta'\left( a_\Psi\atop b_\Psi\right)\bar\eta^{-2}
 \hat{\bar\theta}^2\left( a_I\atop b_I\right)
 \prod\eta^{-1}\theta\left(a_j\atop b_j\right)
 \prod\bar\eta^{-1}\bar\theta\left(a_k\atop b_k\right)
 $$
where  512 is normalizing factor
and products calculated over all fermions excluding fermions
 which $Q$ operators apply to. Namely $\theta'$ and $\hat{\bar\theta}$
 denote action of helicity and gauge group operators respectively.
 The sum is taken over all pairs of boundary condition vectors that
 appear on the model.
According to \cite{Dienes'} we expand the resulting expression with
$\theta$-functions via $q$ and $\bar q$ in order to achieve appropriate
precision.

The final results for Model 1 are as follows ($\delta$ denotes the difference
between corresponding quantities for $U(1)_I$ and $SU(5)_I$).
$\delta b=26.875; \qquad \delta\Delta=5.97 $
Given this relative threshold corrections we can compute its effect on
string unification scale $M_{SU}$.
We find that the correction unification scale is:
\begin{equation}\label{Mstrcor}
M_{SU}^{corr}=M_{SU} \,\exp\frac{({\Delta}_5-{\Delta}_1)}{2(b_5-b_1)}
\approx g_{str.}\cdot 5.6\cdot 10^{17} GeV\ .
\end{equation}

However there are some  ways to explain the difference between
scales of string ($M_{SU}$) and ordinary ($M_{G}=M_{SU}$) unifications
(without additional intermediate exotic vector matter fields
that does not fit into $5$ or $\bar 5$ representations of SU(5).
\cite {Dienes'})
Perhaps the most natural  way is related to  the $G^I \times G^{II}$
String GUT.
If one uses the breaking scheme $G^I\times G^{II}\,\rightarrow G^{sym}$
( where $G^{I,II}=U(5)\times U(3)_H \subset E_8$ ) on the $M_{sym}$-scale,
then unification scale $M_{MSSM}= M_{G} \sim 2\cdot 10^{16}\,$GeV is
the scale of breaking
the $G_{sym}$ group, and string unification do supply the equality
of coupling constant $G\times G$ on the string scale
$M_{SU}\sim g \cdot 5\cdot10^{17}\,$GeV.
Otherwise, we can have an addition scale of the symmetry breaking
$M_{sym} > M_{G}$.
In any case on the scale of breaking
$U(5)^I\times U(5)^{II}\,\rightarrow U(5)^{sym}$
the gauge coupling constants satisfy the equation
\begin{equation}
({\alpha}^{Sym})^{-1}= ({\alpha}^I)^{-1} + ({\alpha}^{II})^{-1} \ .
\label{symscale}
\end{equation}
Thus in this scheme the knowledge of scales $M_{SU}$ and $M_{Sym}$ gives
us a principal possibility to trace the evolution of coupling
constant of the original  group
$SU(5)^I \times U(1)^I \times SU(5)^{II} \times U(1)^{II}$  through
the $SU(5)^{Sym} \times U(1)^{Sym}$ to the low energies
and estimate the values of all coupling constants including
the horizontal gauge constant $g_{3H}$.

The coincidence of $\sin^2 \theta_W$ and ${\alpha}_3$ with experiment will
show how realistic this model is.
The evolution of the gauge constant from the string constant $g_{str}$
to the scale of $M_G$ is described by the equation:
\begin{equation}
({\alpha}_{5,1}^{Sym})^{-1}(M_{G})
=2({\alpha}_{5,1}^{Str})^{-1}(M_{SU}) +
\frac{(b_{5,1}^{I}+ b_{5,1}^{II})}{2 \pi}\ln(M_{SU}/M_{Sym})
+\frac{b_{5,1}^{Sym}}{2 \pi}\ln(M_{Sym}/M_{G})
\label{RGES}
\end{equation}

where ${\alpha}_{5}^{Str}(M_{SU})=g_{str}^2/4 \pi$
and $g_{str}$ is the string coupling.
\bigskip
Now we can get the relation between $g_{str}=g$ and $M_{Sym}$ from RGE's
for gauge running constants $g_5^{Sym}=g_5$, $g_5^{I}$ and , $g_5^{II}$
on the $M_{G}-M_{SU}$ scale.
For example in  Model 1 for the breaking scheme a) one can get:
\begin{equation} \label{b5values}
b_5^{Sym}=12,\,\,\,
b_5^{I}=5,\,\,\,
b_5^{II}=-3
\end{equation}
and
\begin{equation} \label{b1values}
b_1^{Sym}=27,\,\,\,
b_1^{I}=\frac{105}{4},\,\,\,
b_1^{II}=\frac{73}{4}.
\end{equation}

Let us try to make the behaviours of
these coupling constants consistent above and below $M_{G}$ scale.
To do this we have to remember that
there are two possibilities to embed quark-lepton matter in
$SU(5)\times U(1)$ of SO(10) multiplets,
$\underline {1}_{5/2} $,
$ \underline {\bar 5}_{-3/2} $, $\underline {10}_{1/2}$.
For the electromagnetic charge we get:
\begin{eqnarray}\label{eq23}
Q_{em}&=& Q^{II} - Q^{I} = {\bar T}_3 + \frac{1}{2}{\bar y},\nonumber\\
{\bar T}_3&=&T^{II}-T^{I};\,\,\,\,\quad
{\bar y}= y^{II} - y^{I}.
\end{eqnarray}
where
\begin{equation} \label {23'''}
\frac{1}{2}{y^{I,II}}
=\alpha {T^{I,II}}_{5z} + \beta {Y^{I,II}}_5\ , \quad
{T^{I,II}}_{5z} = \mbox{diag} (-1/3,-1/3,-1/3,1/2,1/2).
\end{equation}
In usual non-flipped Georgi-Glashow  SU(5)
model the $Q_{e.m.}$ is expressed
via SU(5) generators only: $\alpha= 1;\,\,\, \beta= 0$. For flipped
$SU(5)\times U(1)$ we have
$\alpha=- \frac{1}{5};\,\,\, \beta= \frac{2}{5}.$
Note, that this charge quantization does not lead to exotic
states with fractional electromagnetic charges (e.g. $Q_{em} =\pm 1/2,
\pm 1/6$)\cite{27',25'}. Also in non-flipped
$SU(5) \times U(1)$ gauge symmetry breaking
scheme  there are  no  SU(3) color triplets and SU(2) doublets with
exotic hypercharges.

For example in flipped (non-flipped) case of Model 1  we can
use the Higgs doublets from
$(5,1;1,1) + ({\bar 5},1;1,1)$,
(the fields $\hat{\phi}_2+\bar{\hat{\phi}}_2$ from sector 5)
for breaking the SM symmetry and low energy $U(1)_5^{sym}$-symmetry.
Below the $M_{G}$  scale in non-horizontal sector
the evolution of gauge coupling constants is described by equations
\begin{eqnarray}
&&\alpha^{-1}_S(\mu )=\alpha^{-1}_5(M_{G} ) + \frac{b_3}{2\pi}
\ln (M_{G}/\mu) \\
&&\alpha^{-1}(\mu )\sin^2\theta_W =\alpha^{-1}_5(M_{G} ) +
\frac{b_2}{2 \pi}\ln (M_{G}/\mu) \\
&&\frac{15k^2}{k^2+24}{\alpha}^{-1}(\mu )\cos^2\theta_W =
{\alpha}_5^{-1}(M_{G}) +\frac{{\bar b}_1}{2 \pi}\ln (M_{G}/\mu)  ,
\end{eqnarray}
where for $N_g=4$ generations and for the minimal set of Higgs fields we have:
$$b_3=-1\ ,\quad b_2=3\ ,\quad
{\bar b}_1=\frac{25k^2}{k^2+24}\cdot b_1 |_{4\ gen.}
=\frac{15k^2}{(k^2+24)}\frac{43}{3} .$$

From these equations and from the experimental data we can find for $N_g=3,4$,
respectively:
\begin{eqnarray}
M_{G}&=&1.2\cdot 10^{16}\mbox{GeV}\ ,\qquad \alpha_5^{-1}=24.4,
\qquad k^2=0.98\ ,\nonumber\\
M_{G}&=&1.17\cdot 10^{16}\mbox{GeV}\ ,\qquad \alpha_5^{-1}=14.1,
\qquad k^2=0.97.
\end{eqnarray}

Here we assume that additional Higgs doublets and triplets appeared
in the theory are heavy ($>M_G$).

From the other hand for the Model 1 and with the mass of the fourth
generation  sufficiently heavy to be invisible but less than $M_G$
the equations
for ${\alpha}_{5,1}^{I.II.Sym}$ for all $M_{sym}$-scale in the range,
$M_G <M_{sym}< M_{SU}$,
give the contradicting value for $k^2$ that is considerably less than 1.
For example, for $M_{Sym}=1.6 \cdot10^{17}GeV$, we get:
\begin{eqnarray}
 M_{SU} = 9.6\cdot10^{17}GeV,\,\,\,
g_{str}=1.7\,\,\rightarrow k^2=0.44.
\end{eqnarray}

In the non-flipped case in Model 1  we have an additional neutral singlet
$\hat{\sigma}^1$ field, which could be used for breaking $U(1)^{Sym}$ group
(of  $U(1)_5^I \times U(1)_5^{II}$) at any high scale, independently on
the $M_{Sym}$ scale, where the
$$SU(5)^I\times SU(5)^{II}\longrightarrow SU(5)^{Sym}$$
 Therefore we have no
constraints on $k^2$ parameter in the range from string scale down to $M_G$.
As a result in  non-flipped case of the
$G^I \times G^{II}$ ($G=SU(5) \times U(1)\times G_H$) GUST the string
unification scale, $M_{SU}$, can be consistent with the
$M_{G}$ ($M_{MSSM}$) scale, i.e using low energy values of the
$g_{1,2,3}$- coupling constants and their RGE (\ref {RGEi}, \ref {bi},
\ref {symscale},\ref {RGES})
 we get for the GUST scale $M_{SU}$  the expression
(\ref {Mstrcor} )
with the corresponding value of the string coupling constant.

In this case while $M_{sym}$ changes in the range from
$M_G=1.17\cdot 10^{16}$ GeV to $10^{18}$ GeV we could
expect the string constant and string scale (\ref{Mstrcor}) to be in the range
$$g\sim (1.40\div 2.13)\ ,
\quad M_{SU}\sim (0.79\div1.20) \times 10^{18} GeV\ . $$

It is interesting to estimate the value of horizontal coupling constant.
The analysis of RG-equations allows to
state that the horizontal coupling constant $g_{3H}$ does not exceed
the electro-weak one $g_2$.

In Model~1 after cancellation of the U(1) anomaly by VEVs
of the fields ${\hat\phi}_1$ or ${\hat\phi}_3$
corresponding $SU(3)^{II}$ or $SU(3)^{I}$ horizontal gauge
symmetry group survives.

Using RG equations for the running constant $g_{3H}^{I,II}$
and the value of the string coupling constant $g_{str}$ at $M_{SU}$
we can estimate a value of
the horizontal coupling constant at low energies.
For Model~1 we have
$$
b_{3H}^{I}= 21\ ,\qquad
b_{3H}^{II}= 13\ ,$$
and
we find from RGE for $g_{3H}$ that
$$ g^I_H\sim 0.3\ ,\qquad g^{II}_H\sim 0.4\ .$$
Below we consider in details the gauge symmetry breaking by VEV's
b), c), d) applied to Model 1 both in flipped and non-flipped cases.
We assume that some additional Higgs doublets and triplets originating
from the group $G^{II}$ (see sector 1 in table \ref{tabl3'}) could
be lighter than $M_G$. The fact that this Higgs fields were initially
(i.e. before breaking $G^I\times G^{II}\rightarrow G$) related to
the group $G^{II}$ excludes their dangerous interaction with matter
that could lead to proton decay.

Also we investigate the dependency on the fourth generation mass $M_4$
and take into account the SUSY threshold.

In general the RGE with thresholds between $M_Z$ and $M_{G}$ are as follows
($k^2\equiv 1$ for non-flipped $SU(5)$):
\begin{eqnarray}\label{rgeeq}
\alpha_3^{-1}&=&{\alpha_5^{-1}} +\frac{1}{2\pi}\ (z + b_3 \ln M_{G}
-b_3^0 \ln M_Z)
\nonumber \\
\alpha_2^{-1}&=&{\alpha_5^{-1}} +\frac{1}{2\pi}\ (y + b_2 \ln
M_{G}-b_2^0 \ln M_Z) \\
\frac{25k^2}{k^2+24}\ \alpha_1^{-1}&=& \alpha_5^{-1}
+\frac{1}{2\pi}\ (x + {\bar b}_1 \ln M_{G}-{\bar b}_1^0 \ln M_Z)\nonumber
\end{eqnarray}
${\bar b}_1,\ b_2, \ b_3$ denote beta function coefficients for the corresponding
coupling constants that take into account all fields below $M_{G}$ scale.
Similarly ${\bar b}^0_1,\ b^0_2,\ b^0_3$ are beta function coefficients that
take all fields of the MSSM excluding superpartners.
We also introduce thresholds factors $x,\  y,\ z$ that depend on various
thresholds $M_I$:
$$ x,y,z= -\sum_I \Delta b^I_{1,2,3}\ \ln M_I $$
In particular we are going to consider SUSY threshold $M_{SUSY}$,
4th generation masses  $M_4$ and effective masses of addition doublets
and triplets $M_{2,3}$.

 In the context of Model 1 we  have
$$ {\alpha^I_5}=\alpha_5(1+q^2),\quad q=g^I_5/g^{II}_5,\quad
{g^I_5}^2=\frac{16\pi^2 g^2_{str}}{16\pi^2 -g^2_{str}b^I_5 \ln\left(\frac{M^2_{G}}{M^2_{str}}
\right)}
$$
and the similar formulae for $g^{II}_5,\ g^I_1,\ g^{II}_1$.

In this scheme the representations like $({\bf 5, 3})$ of the Model 1
break into
the equal number of vector-like triplets and doublets  ( $({\bf 3,1})$ and
$({\bf 1,2})$ under the $SU(3)\times SU(2)$ group). We consider the case
when the fields in one $({\bf 5, 3})$ representation  with the masses below
$M_{G}$ are from the second group $U(5) \times U(3)$.
(In this case we have no problems with the dimension four, five,
six operators of proton decay in Higgs sector.)
 Hence in addition to the MSSM Higgs  vector-like
doublet we have 2 doublets and 3 triplets.

Below are the values of the  $b$ coefficients for our case:
$$
{\bar b}_1=\frac{275k^2}{24+k^2}\ ,\ b_2=5\ ,\ b_3=2\ ,\
{\bar b}^0_1=\frac{105k^2}{24+k^2}\ ,\ b^0_2=-3\ ,\ b^0_3=-7\ .
$$

 Considering  flipped $SU(5)$ case we have to pay attention to the
consistency of the value of $k^2$ derived from (\ref{rgeeq}) and from
RGE of  the string coupling $g_{str}$ above the $M_{G}$ scale.
We use $b$
  coefficients from (\ref{b5values}, \ref{b1values}).

From the system (\ref{rgeeq}) we can get a set of appropriate masses
in the range of $M_Z-M_{G}$ and values of $\alpha_5,\ k^2$ and $q^2$
as well. But then we should apply RGE between the string scale
and $M_G$ scale
to check out whether
this values are consistent. This equations give us $k^2 < 1$.
 Our calculations show
that with  $k^2\leq 1$ for flipped $SU(5)$ in the Model 1
 one cannot get appropriate values for
$M_{SUSY},\, M_{2,3,4},\,M_{G}$ (i.e. that are within range
$M_{SUSY}-M_{G}$) that are consistent with string RGE.

For non-flipped case we apparently obtain the demand that
constants $\alpha_{1,2,3}$ converge to one point (that is equivalent
to $k^2\equiv 1$) which is consistent with RGE in the framework
of the MSSM-like models.

For this case we consider the b)  breaking way of
$ [SU(5) \times U(1)]^{\otimes 2}$. We can consider the cases c) and d)
as a limits ($x \ll y$ and $x \gg y$). As it follows from our analysis
there exists a range of parameters values (threshold masses) that make
system (\ref{rgeeq}) consistent and we have an appropriate hierarchy
of the scales.

The maximum value of $M_G$ one can obtain in this case is
$M_G\sim 1.3\cdot 10^{16}$ GeV. The mutual dependencies of the threshold masses
are shown on Fig. 1 where $M_{2,3}$ are effective masses because the equation
(\ref{rgeeq}) depends on them only ($M_2=\sqrt{M_2^{(1)} M_2^{(2)}}$;
$M_3=\sqrt[3]{M_2^{(1)} M_2^{(2)} M_2^{(3)}}$).

Note that the Higgs triplets and doublets considered obtain their masses
via $F^2$-term of the field ${\hat\Phi}_{(5,1;1,3)}$ (see the first term
in the $W_3$ (\ref{w3})). Hence $M^2_3\sim |x|^2$; $M^2_2\sim |y|^2$.
At the same time  squared masses of the vector bosons of the broken
groups $SU(3)$ and $SU(2)$ are proportional to $(g^2_I+g^2_{II})|x|^2$;
$(g^2_I+g^2_{II})|y|^2$. This means that  in general above the thresholds
$M_{2,3}$ we should take into account the restoration of the symmetry
$SU(n)\longrightarrow SU(n)_I\times SU(n)_{II},\ \ \ n=2,3$.
I.e. our plots are correct only in the region with $M_2$ close to $M_3$.
The other cases demand more careful accounting of the symmetry
restoration thresholds. This question is currently under
consideration and we will present the results in the future.

The horizontal gauge constant on the scale of 1 TeV for first
or second $SU(3)_H$ group (depending on which of them will survive
after suppression of the $U(1)$ anomaly) appears to be of the order

$${g_{3H}^I}^2 \biggl( O(1\mbox{TeV})\biggr)\approx 0.10\div 0.11\ , \quad
{g_{3H}^{II}}^2 \biggl( O(1\mbox{TeV})\biggr)\approx 0.15\div 0.17\ .
$$

The calculations for our model for different breaking chains
show that for evaluation of intensity
of a processes with a gauge horizontal bosons at low energies
we can use inequality
${{\alpha}_{3H}(\mu)}\,\leq \,{{\alpha}_2(\mu)}\ .$

This work was supported in part by Russian Foundation for Basic Researches,
grant No 95-02-06121a. One of us (G.G.V.) would like to thank INFN
for financial support, especially Prof. G. Puglierin and Prof. C. Voci,
for hospitality during his stay in Padova. It is a great pleasure for him
to thank
Profs. G. Costa, and A. Masiero for useful discussions, for help and support.
Many  helpful advises and  interesting discussions with R. Barbieri,
J. Ellis, H. Dahmen, K. Dienes, J. Gasser,
L. Ibanez, H. Leutwyler, P. Minkowsky, N.Paver, G.Ross and F. Zwirner during
this work are also acknowledged.

\begin{figure}[h]
\label{grafik2}
\vskip -7cm
\vbox to 23cm {\hbox to 14cm {\epsfxsize=14cm\epsffile{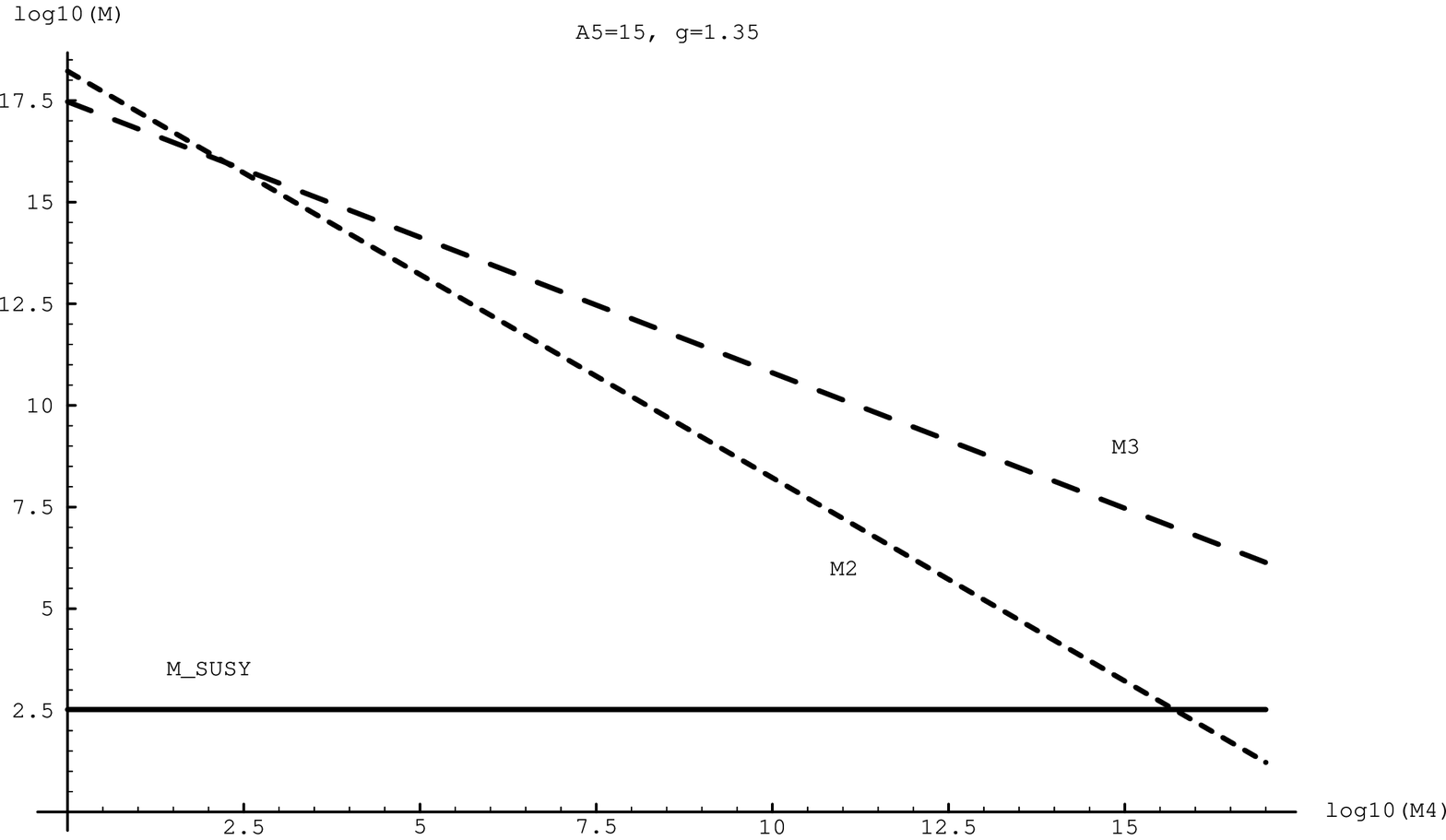}}%
\vskip -9cm
\hbox to 14cm {\epsfxsize=14cm\epsffile{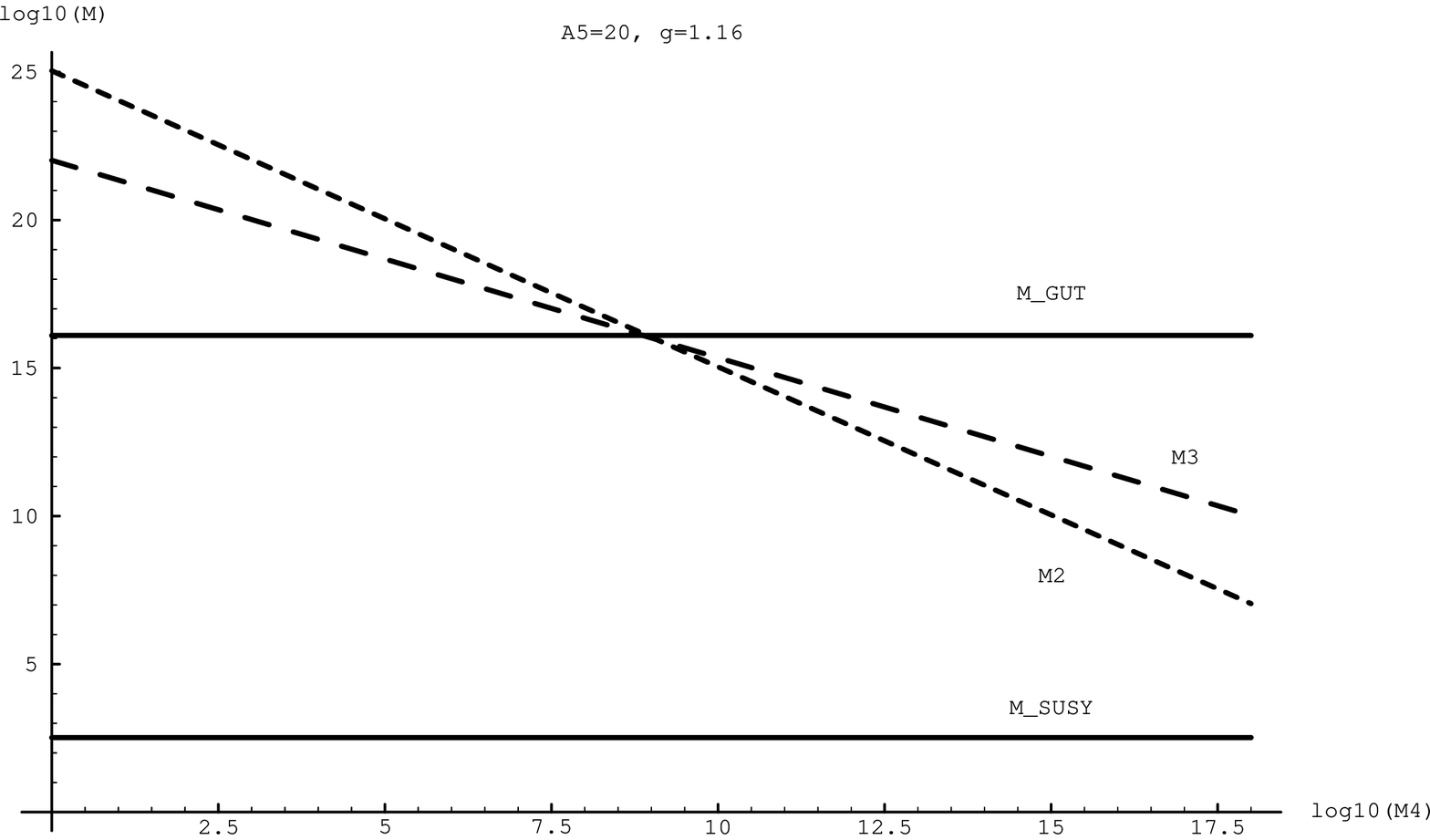}}}
\caption{$M_{2,3}$ behavior.
 {\tt A5}$=\alpha_5^{-1}$, $M_G=1.26\cdot 10^{16}$, {\tt g} means
 $g_{str}$, $g^I_3=0.32,\ \ g_3^{II}=0.40$.}
\end{figure}

\end{document}